\documentclass{sig-alternate}
\newcommand{\ignore}[1]{}
\usepackage[pass]{geometry}
\usepackage{fancyhdr}
\usepackage[normalem]{ulem}
\usepackage[hyphens]{url}
\usepackage{hyperref}
\usepackage{color}
\usepackage{soul}
\usepackage[algoruled,boxed,lined]{algorithm2e}
\usepackage{listings}
\usepackage{multirow}
\usepackage{caption}
\usepackage{makecell}
\usepackage{tablefootnote}
\usepackage[para,online,flushleft]{threeparttable}
\usepackage{enumitem}

\newcolumntype{C}[1]{>{\centering\let\newline\\\arraybackslash\hspace{1pt}}m{#1}}
\newcolumntype{L}[1]{>{\raggedright\let\newline\\\arraybackslash\hspace{1pt}}m{#1}}

\newcolumntype{R}[1]{>{\raggedleft\let\newline\\\arraybackslash\hspace{1pt}}m{#1}}

\usepackage[keeplastbox]{flushend}
\newcommand{\algrule}[1][.5pt]{\par\vskip.2\baselineskip\hrule height #1\par\vskip.2\baselineskip}

\newcommand{\hpcasubmissionnumber}{302}

\fancypagestyle{firstpage}{
  \fancyhf{}
\setlength{\headheight}{50pt}

  \fancyhead[C]{\normalsize{HPCA 2020 Submission
      \textbf{\#\hpcasubmissionnumber} -- Confidential Draft -- Do NOT Distribute!!}}
  \pagenumbering{arabic}
}

\title{Leaking Information Through Cache LRU States\vspace{-36pt}} 

\begin{document}

\numberofauthors{2} 
\author{
\alignauthor
Wenjie Xiong\\
	\affaddr{Yale University}\\
       \affaddr{New Haven, CT 06520, USA}\\
       \email{wenjie.xiong@yale.edu}
\alignauthor
Jakub Szefer\\
       \affaddr{Yale University}\\
       \affaddr{New Haven, CT 06520, USA}\\
       \email{jakub.szefer@yale.edu}
}

\maketitle

\pagestyle{plain}

\begin{abstract}

The Least-Recently Used cache replacement policy and its variants
are widely deployed in modern processors.
This paper shows for the first time in detail that the LRU states of caches can be used to leak information:
any access to a cache by a sender will modify the LRU state, and the receiver is able to observe this through a timing measurement.
This paper presents LRU timing-based channels both when the sender and the receiver have shared memory, e.g., shared library data pages,
and when they are separate processes without shared memory.
In addition, the new LRU timing-based channels are demonstrated on both Intel and AMD processors
in scenarios where the sender and the receiver
are sharing the cache in both hyper-threaded setting and time-sliced setting. 
The transmission rate of the LRU channels can be up to 600Kbps per cache set in the hyper-threaded setting.
Different from the majority of existing cache channels which require the sender to trigger cache misses, the new LRU channels work with the sender only having cache hits, making the channel faster and more stealthy.
This paper also demonstrates that
the new LRU channels can be used in transient execution attacks, e.g., Spectre. 
Further, this paper shows that
the LRU channels pose threats to existing secure cache designs, and
this work demonstrates the LRU channels affect the secure PL cache.
The paper finishes by discussing and evaluating possible defenses.

\end{abstract}


\section{Introduction}
\label{sec:intro}

Side channels and covert channels in processors have been gaining renewed attention in recent years~\cite{szefer2018survey}.
Many of these channels leverage the timing information.
To date, researchers have shown numerous timing-based channels in caches,
e.g.,~\cite{yarom2014flush, osvik2006cache,bonneau2006cache,liu2015last,yan2019attack,yao2018coherence},
as well as other parts of the processor, such as the shared functional units in simultaneous multithreading (SMT) processors,
e.g.,~\cite{wang2006covert,yarom2017cachebleed,schwarz2018netspectre,moghimi2018memjam,aldayaport,bhattacharyya2019smotherspectre,evtyushkin2018branchscope,evtyushkin2016jump}.
The canonical example of timing channels are the channels in caches,
where timing reveals information about cache states.
This in turn can be used to leak information, such as cryptographic
keys, e.g.,~\cite{gullasch2011cache,percival2005cache,bernstein2005cache, bonneau2006cache,aciiccmez2006trace}.
Further, many of the variants of the recent Spectre and Meltdown attacks also use covert channels, in addition to transient execution, to exfiltrate 
data, e.g.,~\cite{kocher2018spectre,lipp2018meltdown,canella2018systematic}.

In processor caches, the order in which the cache lines are evicted
depends on the cache replacement policy.  Normally, different variants of the Least-Recently Used (LRU) policy are
implemented in modern processors, such as Tree-PLRU~\cite{so1988cache} or Bit-PLRU~\cite{malamy1994methods}.
In a cache, the LRU state is maintained for each cache set,
and it is used to determine which cache line in the cache set should be evicted
when there is a cache miss causing a cache replacement.
The LRU state is updated on every cache accesses to indicate which cache line in the set
was just accessed.
Thus, both cache hits and misses in the set cause updates to the LRU state of the set.

The basis of the new LRU timing-based channels is the timing of the cache accesses, as it is affected by the LRU states.
Thus, 
the LRU channels work even when the sender only triggers a cache hit,
and the receiver later triggers a possible replacement and then measures the time -- unlike prior attacks, which require a cache miss to be triggered by the sender.
This makes the attacks more stealthy.  It may also allow the attacks to bypass defenses such as based on performance counters~\cite{nomani2015predicting}
where behavior of cache missies is monitored.
Moreover, lack of required missies for the sender benefits the transient execution attacks, as only a small speculation window is required for the sender to trigger a cache hit, compared to a miss.

The new LRU timing-based channels are also a threat to many of the existing secure caches proposals.
Numerous secure caches~\cite{lee2005architecture,wang2007new,domnitser2012non,zhang2012language,yan2017secure,kiriansky2018dawg,liu2014random,liu2016newcache,keramidas2008non,yan2018invisispec,khasawneh2018safespec} have been presented, and they aim to either partition or randomize
the victim's and the attacker's  cache accesses to defend the cache timing-based side channels.
However, most of the secure caches have not considered the LRU states and are vulnerable to the new LRU channel.
Especially, this paper demonstrates the vulnerability to the new LRU-based attacks in 
the well-known Partition-Locked (PL) cache~\cite{wang2007new}, and then shows how to mitigate the attacks in the PL cache. 

In this paper, the new LRU timing-based channels are demonstrated and evaluated in-depth for the first time.
The biggest challenge of the LRU channels is how the receiver can accurately observe which level of cache a
memory access hits in, i.e. how to measure the timing precisely. 
This paper proposes to use dedicated data structures and a pointer chasing algorithm
in the receiver's program to allow for fine-grained measurements of the latency of the memory accesses.
Further, two algorithms are designed to build LRU timing channels: both with and without shared
memory between the sender and the receiver, making the LRU channels practical in a variety of attack scenarios.
We evaluated the LRU channels on a number of commercial processors
including Intel and AMD processors with different microarchitectures, and both hyper-threaded and time-sliced sharing settings are considered.
The LRU channels are also demonstrated in  Spectre attack.
The contributions of this work are as follows:

\vspace{-0.5em}
\begin{itemize}[itemsep=-0.2em]
\item The first detailed presentation of how the LRU states in caches can be used as a timing-based
side and covert channels for information leaks, both with and without shared memory between the\\sender and the receiver.
\item Detailed analysis and evaluation of the LRU channels, including
evaluation of the transmission rates and bit error rates of the LRU covert channels on both Intel and AMD processors and comparison of the LRU channels with the existing cache channels from the perspective of encoding time and cache miss rates.
\item Demonstration of LRU channels in transient execution attacks.
\item Demonstration in {\tt gem5} simulator of how the LRU channels break the security of PL cache~\cite{wang2007new}, and how it can be fixed.
\item Proposal for, and evaluation of, mitigations of the LRU channels.
\end{itemize}

\section{Background}
\label{sec:background}

\subsection{Timing-Based Cache Channels}

There are typically two types of  timing-based cache side and covert channels. One type leverages the contention in the cache bank~\cite{yarom2017cachebleed,moghimi2018memjam}. 
The other leverage the states in the cache, e.g., tag state (if a certain address is in the cache)~\cite{yarom2014flush, osvik2006cache,bonneau2006cache,liu2015last} or cache coherence state~\cite {yao2018coherence}. 

Like other side and covert channels leveraging port contention~\cite{wang2006covert,schwarz2018netspectre,aldayaport}, channels leveraging the contention in the cache bank~\cite{yarom2017cachebleed,moghimi2018memjam} 
require the sender and the receiver to execute concurrently as two hyper-threads.

Channels using cache states leverage the fact that whether a cache line is available in the cache or not affects the timing of the cache operations.
The sender and the receiver do not have to be two concurrent hyper-threads. They can be within one thread or share the cache in time-sliced setting.
All these existing channels, however, require a cache miss by the sender  
to change the cache state when the sender is sending information. For example, in Flush+Reload attacks~\cite{yarom2014flush}, the sender will need to access the cache line that was previously flushed to memory by the receiver. 
Thus, the access will cause a cache miss.
Meanwhile, any cache access, both cache hit or miss, can trigger the new LRU attack.

\subsection{Cache Replacement Policy}
\label{sec:background_LRU}

When a cache line is accessed but it is not in the cache (i.e., a cache miss), the cache line will be fetched into the cache set. In this
case, another cache line needs to be evicted from the cache set to make room for the incoming cache line.
The replacement policy selects a cache way from the set to evict, known as the {\em victim way}.
The replacement algorithm uses some state to store the history of accesses to cache ways in a given set.
In L1 cache, the LRU policy and its variants are most widely
used because they give high cache hit rate.
In last level cache (LLC), due to the reduced data locality, other replacement policies can be used~\cite{jaleel2010high,qureshi2007adaptive}.

\textbf{LRU:} The LRU algorithm keeps track of the age of cache lines. 
If a cache replacement is needed on a cache miss,
the least recently used cache way (i.e., oldest way) will be selected to be the victim way and will be evicted.
In an $N$-way cache, $log(N)$ bits are used per cache line per way to store the age of the line,  for a total of $N log(N)$ bits for each cache set.
The ``true'' LRU algorithm is expensive in terms of latency (to update LRU states) and area
(to store the age of all the cache lines).
So often a variant of a Pseudo Least-Recently Used (PLRU) is used instead.

\textbf{Tree-PLRU:} The Tree-PLRU~\cite{so1988cache} policy uses a binary tree structure to keep track of the cache access history in a cache set.
Each tree node indicates whether the left sub-tree or the right sub-tree has been less recently used.
To find the victim way, the replacement algorithm starts from the root and always goes to the less recently used child
to find the leaf node that indicates the victim way.
To update the Tree-PLRU when a cache line in a way is accessed, all the nodes on the path from the root to the accessed
way's leaf node are set to point to the child that is not the ancestor of the accessed cache way.
For an $N$-way cache, the tree has $N-1$ nodes with each taking 1~bit, for a total of $N-1$ bits for each cache set.

\textbf{Bit-PLRU:} The Bit-PLRU~\cite{malamy1994methods} policy, which is also called Most Recently Used (MRU) policy,
uses one bit to store the history of each cache way, called {\em MRU-bit}.
When a way is accessed, its MRU-bit will be set to 1, indicating the way is recently used.
Once all the ways have the MRU-bit set to 1, all the MRU-bits are reset to 0.
To find a victim, the way with the lowest index whose MRU-bit is 0 is chosen.
For an $N$-way cache set, a total of $N$ bits are required.
The logic of the Bit-PLRU is simpler than Tree-PLRU.

\section{Threat Model and Assumptions}
\label{sec:threat_model}

\begin{figure}[t]
\centering
\includegraphics[width=3in]{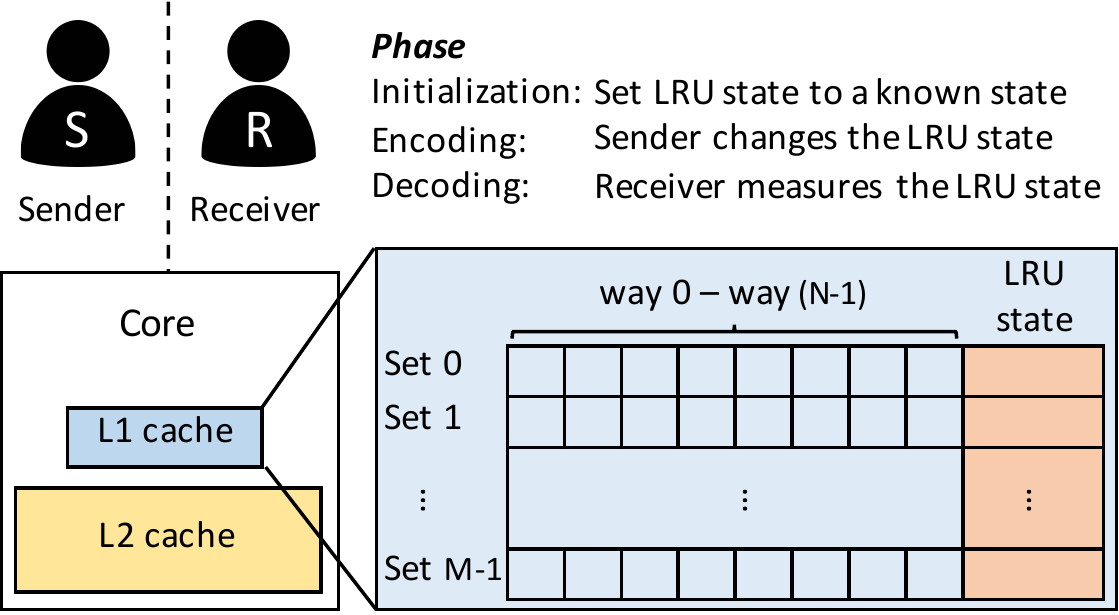}
\caption{\small Cache organization and the phases of the new LRU timing-based side and covert channels.}
\vspace{-6pt}
\label{fig:overview}
\end{figure}

In this paper, we demonstrate a covert channel, and we always use the term sender and receiver. A cache covert channel can be extended 
to a side channel when the victim has secret-dependent accesses~\cite{yarom2014flush, osvik2006cache,bonneau2006cache,liu2015last,yarom2017cachebleed}. 

We assume $N$-way set-associative caches and further assume the cache uses an LRU, Tree-PLRU, or bit-PLRU replacement algorithm which evicts the least recently used cache line.
Like all other side or covert channels, the LRU timing-based channel involves two parties: the sender and the receiver.
Following techniques used in \cite{ristenpart2009hey,zhang2014cross}, we assume the two parties can be co-located on the same core to share the L1 cache, as shown in Figure~\ref{fig:overview}, 
either in an SMT machine as two hyper-threads running in parallel or as two threads time-sharing the core.
The LRU states of the shared cache can be influenced (by the sender) and observed (by the receiver).
Existing attacks, such as side 
channels~\cite{moghimi2018memjam,aldayaport,bhattacharyya2019smotherspectre,evtyushkin2018branchscope,evtyushkin2016jump}
or Spectre attacks using Branch Target Buffer (BTB) or  Return Stack Buffer (RSB)~\cite{kocher2018spectre,koruyeh2018spectre,maisuradze2018ret2spec},
show that sharing of the same physical core 
is practical and poses real threats for computer systems.

In this paper, we focus on the LRU states in L1 cache. 
The LRU channels in the other levels of caches are also possible\footnote{Concurrently to this submission,
a preprint paper~\cite{briongos2019reload} has been recently posed on {\em arXiv} on side channels that leverage
the replacement policy in LLC. However, our work demonstrates the LRU channels both with and without shared memory and  without using {\em clflush} instruction.}. 
Depending on the cache architecture, for the sender to update the LRU states of the lower level of caches, a miss in the higher cache level is required, 
e.g., the sender's accesses to L1 or L2 caches will not change the replacement state in the LLC. Especially, L1 is directly accessed by the processor pipeline and L1 LRU state is updated on every memory access.
Thus, attack  in the LRU states of L1 is more stealthy. And the timing channels in LRU states in L2 or LLC can be detected or protected by the existing cache side channel detection or protection techniques in L1 and prefetching the secure-relevant data to L1. 

For all types of attacks, we assume the receiver can extract useful information from  
the memory access pattern of the sender, which modifies the LRU states.

\section{LRU Timing-based Channels}
\label{sec:LRU_channel}

Our new LRU timing-based channels leverage the LRU states of cache sets.
In this section, we discuss how the LRU state in {\em one} cache set can be used to transfer information,
which is referred to as the {\em target set}.

The LRU state for each set contains several bits,
thus it is possible to transfer more than 1 bit per target set.
However, limited by the fact that any access to the set will change the LRU state,
we focus on letting the receiver only measure the set once.
Especially, the receiver can observe the timing of one memory access which can only have two results: a cache hit or a cache miss.
Thus, at most one bit can be transferred per cache set at one time.
To transfer information using an LRU channel, in general, there are three phases:

\textbf{Initialization Phase:}
First, a sequence of memory accesses is performed so that the LRU state is partially known to the receiver.

\textbf{Encoding Phase:}
To send information, the sender accesses one or more memory locations mapping to the target set to change the LRU state.
The pattern of memory accesses depends on the information to be sent.
Algorithms in this paper are designed to be light-weight in the encoding phase, where the sender only needs to do at most one memory access.

\textbf{Decoding Phase:}
The receiver first accesses one or more memory locations mapping to the target set to
potentially trigger a cache replacement and cause a cache line to be evicted based on the LRU state.
The receiver then observes the timing of accessing the memory location to learn if the cache line is evicted and thus infer what the LRU state was.

\subsection{LRU Channel with Shared Memory}
\label{sec:Protocol1}

Algorithm~\ref{alg:Protocol1} shows a communication protocol using the LRU cache states assuming shared memory.  
The sender and the receiver first agree on the target cache set they will use to transfer information.
We use the term {\em line $0$--$N$} to denote $N{+}1$ different cache lines that map to the target set. 
This can be achieved by using data in $N{+}1$ different {\em physical addresses} with the same cache index bits but different tag bits.
Note that {\em line $n$} (where $n\in[0,N]$) 
refers to a cache line with a certain physical address and not a specific cache entry,
and the name does not imply certain literal physical address~$n$.
The {\em line $n$} could be placed in any cache way in the set.

In Algorithm~\ref{alg:Protocol1}, the sender and the receiver both need to use the same physical address (or a physical address within the cache line) to access cache line $0$ in the cache.
This can be achieved by a memory location in a shared dynamic linked library, as in~\cite{yarom2014flush}.
Further, $m$ is a 1-bit message to be sent, and $d$ is a parameter indicating how the receiver's accesses are split between the initialization and decoding phase.  
Then, the sender and the receiver can build a channel following Algorithm~\ref{alg:Protocol1}.

For example, when $N=8$ and $d=8$, the sequence of memory accesses when sending $m=0$ is as follows:
\vspace{-0.5em}
\begin{itemize}[itemindent=0pt,leftmargin=15pt,itemsep=-0.3em]
\item Init. Phase: $ 0\rightarrow1\rightarrow2\rightarrow 3\rightarrow4\rightarrow5\rightarrow6\rightarrow7$
\item Encoding Phase: no access
\item Decoding Phase: $ 8\rightarrow0$ (miss)
\end{itemize}
\vspace{-0.5em}
In this 8-way set associative cache, line 0 will be chosen by the LRU policy as the victim way
and will be evicted from L1 when accessing line 8\footnote{With
PLRU replacement algorithms, line 0  is not guaranteed to be evicted. However, as will be evaluated in Section~\ref{sec:PLRU_analysis}, line 0 will be evicted in most of the cases.}, and the receiver will observe L1 miss when accessing line 0 in the end.

Meanwhile, the sequence of memory accesses when sending $m=1$ is as follows:
\vspace{-0.5em}
\begin{itemize}[itemindent=0pt,leftmargin=15pt,itemsep=-0.3em]
\item Init. Phase: $ 0\rightarrow1\rightarrow2\rightarrow 3\rightarrow4\rightarrow5\rightarrow6\rightarrow7$
\item Encoding Phase: ${0}$ (hit)
\item Decoding Phase: $8\rightarrow0$ (hit)
\end{itemize}
\vspace{-0.5em}
During the encoding phase, the access to line $0$ will make it 
become the {\em newest} line in the LRU state, and the remaining accesses in the decoding phase will not evict it.
When the receiver measures the time of accessing line $0$ in the decoding phase,
the receiver will observe an L1 cache hit, and the receiver can infer that the sender has sent $m=1$.

Comparing Algorithm~\ref{alg:Protocol1} with Flush+Reload attack~\cite{yarom2014flush}, both require shared memory, but
the LRU channel does not require explicit flush, and line~0 might always be in the cache, i.e., the sender might only have cache hits.

\begin{algorithm}[t]
\small{
 line $0$--$N$: cache lines mapping to the target set\\
 m: a 1-bit message to transfer on the channel\\
 d: a parameter of the receiver\\
  \algrule
 \textbf{Receiver Operations:}
 \algrule
   // Step 0: Initialization Phase\\
 \For{$i = 0;\ i < d;\ i = i + 1$}{
 Access line i\;
 }
 sleep; //  To allow the sender code to run here for encoding\\
 // Step 2: Decoding Phase\\
 \For{$i = d;\ i < N+1;\ i = i + 1$}{
 Access line i\;
 }
 Access line 0 and time the access\;
 \algrule
 \textbf{Sender Operations:}
 \algrule
  // Step 1: Encoding Phase\\
   \eIf{m=1}{
   Access line 0\;
   }{
   Do not access line 0\;
  }
  }
 \caption{ LRU Channel with Shared Memory}
 \label{alg:Protocol1}
\end{algorithm}

\subsection{LRU Channel without Shared Memory}
\label{sec:Protocol2}

\begin{algorithm}[t]
\small{
line $0$--$N$: cache lines mapping to the target set\\
 m: a 1-bit message to transfer on the channel\\
 d: a parameter of the receiver\\
  \algrule
 \textbf{Receiver Operations:}
 \algrule
    // Step 0: Initialization Phase\\
 \For{$i = 0;\ i < d;\ i = i + 1$}{
 Access line i\;
 }
 sleep; // To allow the sender code to run here for encoding\\
  // Step 2: Decoding Phase\\
 \For{$i = d;\ i < N;\ i = i + 1$}{
 Access line i\;
 }
 Access line 0 and time the access\;
 \algrule
 \textbf{Sender Operations:}
 \algrule
   // Step 1: Encoding Phase\\
   \eIf{m=1}{
   Access line $N$\;
   }{
   Do not access target set\;
  }
  }
 \caption{LRU Channel w/o Shared Memory}
 \label{alg:Protocol2}
\end{algorithm}

In Algorithm~\ref{alg:Protocol2}, the sender and the receiver do not need to access any shared memory location.
The sender and the receiver can map memory accesses to the target set by
using proper virtual memory addresses in their own memory spaces.
For performance, L1 cache is usually virtual-indexed and physical-tagged (VIPT).
For example, for an L1 cache with 64 sets with a cache line size of 64 bytes, bits 6--11 of the address decide the cache set.
The receiver can make sure lines 0--$(N{-}1)$ map to the same set as line $N$ by using memory
locations with bits 6--11 of the virtual address to be the same as line $N$.
Then, the sender and the receiver can build a channel following Algorithm~\ref{alg:Protocol2}.

For example, when $N=8$ and $d=4$, the order of memory accesses when sending $m=0$ is as follows:
\begin{itemize}[itemindent=0pt,leftmargin=15pt,itemsep=-0.2em]
\vspace{-0.5em}
\item Init. Phase: $ 0\rightarrow1\rightarrow2\rightarrow 3$
\item Encoding Phase: no access
\item Decoding Phase: $4\rightarrow5\rightarrow6\rightarrow7\rightarrow0$ (hit)
\end{itemize}
\vspace{-0.5em}
The order of memory accesses when sending $m=1$ is:
\vspace{-0.5em}
\begin{itemize}[itemindent=0pt,leftmargin=15pt,itemsep=-0.2em]
\item Init. Phase: $0\rightarrow1\rightarrow2\rightarrow 3$
\item Encoding Phase: ${8}$ (hit, if line 8 is in cache before Init. Phase)
\item Decoding Phase: $4\rightarrow5\rightarrow6\rightarrow7\rightarrow0$ (miss)
\end{itemize}
\vspace{-0.5em}
Whether the sender accesses line~$8$ or not  will change the LRU state,
and in the decoding phase, it will decide which line will be evicted if the sender's access to line~$7$ misses in the cache.
The receiver will observe an L1 cache hit  when accessing line $0$ if the sender is sending $m=0$, and will observe an L1 cache
miss if the sender is sending $m=1$.
Compared to Algorithm~\ref{alg:Protocol1}, there will be more noise in this channel, as any thread accessing the target set can cause
line $0$ to be evicted. A miss of line~$0$ does not necessarily mean that the sender accessed line~$8$.
The noise is due to no shared memory, and other known cache side channel attacks (e.g., Prime+Probe channel~\cite{osvik2006cache}) also have this source of noise.

Comparing Algorithm~\ref{alg:Protocol2} with Flush+Reload attack, no shared memory is required.
Comparing Algorithm~\ref{alg:Protocol2} with Prime+Probe attack~\cite{osvik2006cache},
in Prime+Probe, the receiver will access the whole set in both the prime and the probe phases, 
and the sender will have a miss between the two phases. Meanwhile, in Algorithm~\ref{alg:Protocol2}, 
the receiver does not access the whole set in either phase. The receiver only needs to measure the time of one memory access in 
LRU channel rather than the time of $N$ memory accesses in the Prime+Probe attack. Moreover, the sender's data line N might always be in the cache.

\subsection{PLRU vs. LRU Replacement Policy}
\label{sec:PLRU_analysis}

In true LRU, the least recently used way is always chosen as the victim.
Consider the following two memory accesses sequences in an 8-way cache,
with each number representing accessing a cache line in the set:
\vspace{-0.5em}
\begin{itemize}[itemindent=0pt,leftmargin=15pt,itemsep=-0.2em]
\item Sequence 1 (access in order):
$0\rightarrow1\rightarrow2\rightarrow3\rightarrow4\rightarrow5\rightarrow6\rightarrow7\rightarrow8$.

\item Sequence 2 (access in order with random insertion):
$0$ $(\rightarrow x)\rightarrow1$ $(\rightarrow x)\rightarrow2$ $(\rightarrow x)\rightarrow3$ $(\rightarrow x)\rightarrow4$ $(\rightarrow x)\rightarrow5$ $(\rightarrow x)\rightarrow6$ $(\rightarrow x)\rightarrow7$. Here, line $x$ is a cache line that maps to this cache set and
is different from lines $0$--$7$. The parentheses indicate the access might happen or not,
and we assume line $x$ will be accessed at least once.

\end{itemize}
\vspace{-0.5em}
If true LRU is used, line $0$ will be evicted in both sequences.
However, in PLRU, line $0$ is not guaranteed to be evicted. Because PLRU uses fewer state bits to track the memory access history,
the cache LRU state before the access sequence could still affect the choice of victim way,
and longer history should be considered when analyzing the PLRU.
Consider the following {\em initial conditions} before accessing the above sequence:
\vspace{-0.5em}
\begin{itemize}[itemindent=0pt,leftmargin=15pt,itemsep=-0.2em]
\item {Random:}  The cache contains some of the lines $0$--$7$ and probably other lines,
and the initial access order of lines $0$--$7$ is~random (e.g., the lines in the set are accessed in a random order probably with lines other than lines $0$--$7$ ).
\item {Sequential:} The cache contains some of the lines $0$--$7$ and probably other lines,
and the initial access of lines $0$--$7$ is in sequential order
(e.g., the set is accessed in order with the random insertion of lines other than lines $0$--$7$ like Sequence~2).
\end{itemize}
\vspace{-0.5em}

\begin{table}[t]
 \caption{\small Probability of line $0$ being evicted with PLRU.}
 \label{table:PLRU_analysis}
\centering
\small
 \begin{tabular}{|c|c|c|c|c|c|c|}
 \hline
\multirow{3}{0.27in}{\textbf{Init. Cond.}} & \multirow{3}{0.27in}{\textbf{Num. Loop Iter.}}  & \textbf{LRU}&\multicolumn{2}{c|}{\textbf{Tree-PLRU}}& \multicolumn{2}{c|}{\textbf{Bit-PLRU}} \\
\cline{4-7}
&&\multirow{1}{0.2in}{\textbf{Seq.\\ 1\&2}}& \multirow{1}{0.2in}{\textbf{Seq. 1}} & \multirow{1}{0.2in}{\textbf{Seq. 2}} & \multirow{1}{0.2in}{\textbf{Seq. 1}} & \multirow{1}{0.2in}{\textbf{Seq. 2}}  \\ 
&&& &  &  &   \\ 
\hline
\multirow{4}{0.1in}{\rotatebox{90}{Random}} &$1$& 100\%& 50.4\% & 62.7\% & 38.5\% & 55.5\% \\
				&$2$& 100\%&  82.8\%& 65.6\% & 55.6\% & 69.7\% \\
				&$3$& 100\%&  99.2\%& 64.2\% & 67.3\% & 80.1\% \\	
				&$>=8$& 100\%&  100\%& $\sim$62\% & 100\% & $\sim$99\% \\
\hline
\multirow{4}{0.1in}{\rotatebox{90}{Sequential}} &1& 100\%& 90.9\% & 75.6\% & 60.4\% & 61.0\% \\
				&2& 100\%&100\% & 65.9\% & 63.0\% & 64.1\%\\
				&3& 100\%& 100\% & 64.0\% & 67.3\% & 70.3\% \\
				&$>=8$& 100\%& 100\%& $\sim$62\%  &100\% & $\sim$99\%\\[0.03in]
 \hline
 \end{tabular}
\end{table}

\begin{table}
\captionof{table}{\small Latency of cache access (cycles).}
 \label{table:cache_latency}
\centering
\small
 \begin{tabular}{|C{1.15in}|p{0.2in}|p{0.15in}|}
 \hline
\textbf{Microarchitecture}  &  \textbf{L1D} & \textbf{L2} \\
\hline
Intel Sandy Bridge & 4-5 & 12\\
 \hline
Intel Skylake & 4-5 & 12 \\
 \hline
AMD Zen & 4-5 & 17 \\
 \hline
 \end{tabular}
\end{table}

\begin{figure}[t]
\centering
\begin{minipage}{0.45\textwidth}
\centering
\small
\begin{lstlisting}[basicstyle=\scriptsize]
rdtscp
movl %eax, %esi  
movq (%rbx),  %rax //L1 hit
movq (%rax),  %rax //L1 hit
movq (%rax),  %rax //L1 hit
movq (%rax),  %rax //L1 hit
movq (%rax),  %rax //L1 hit
movq (%rax),  %rax //L1 hit
movq (%rax),  %rax //L1 hit
movq (%rax),  %rax //target address to measure
rdtscp
subl %esi, %eax
\end{lstlisting}
\captionof{figure}{\small Pointer chasing algorithm used to measure time.}
\label{code:pointer_chasing}
\centering
\end{minipage}
\end{figure}

We implemented an in-house simulator to simulate the Tree-PLRU~\cite{so1988cache} and Bit-PLRU~\cite{malamy1994methods} replacement policies in an 8-way set.
First, in the warm-up phases, we create accesses to the set for each of the possible initial conditions.
Then, Sequence~1 or Sequence~2 is accessed in a loop, and
whether line $0$ is in the cache after each sequence is recorded for each loop iteration.
We repeat the above test in the simulator for $10,000$ times for each configuration,
and present results in Table~\ref{table:PLRU_analysis}.

As shown in Table~\ref{table:PLRU_analysis}, under random initial condition, line 0 might still be kept in the cache with a high probability.
Meanwhile, sequential initial condition gives  a high probability of line 0 being evicted after several loop iterations,
especially for sequence 1 and the Bit-PLRU.
Note that true LRU will always evict line~0.
Thus, to build a covert channel through the LRU states under PLRU policy, the receiver should ensure the sequential initial condition by placing line $1$--$7$
in the receiver's address space and then always accessing them in~order to maximize the success rate.

\subsection{Challenge: Measuring the Latency of L1 Hit and Miss}
\label{sec:measure_L1}
\label{sec:pointer_chasing}

\begin{figure}[t]
\begin{minipage}{0.23\textwidth}
\centering
\includegraphics[width=1.6in]{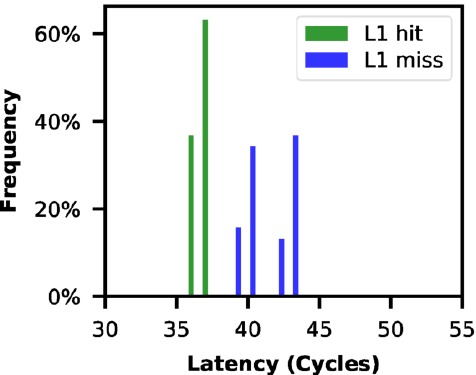}
\end{minipage}
\begin{minipage}{0.23\textwidth}
\centering
\includegraphics[width=1.6in]{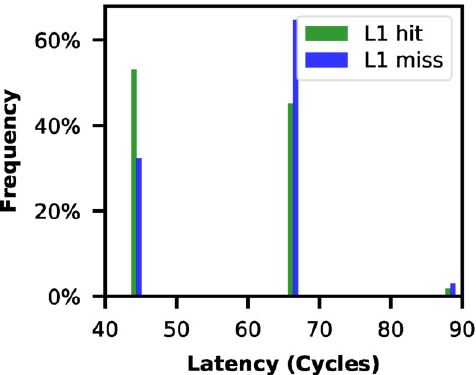}
\end{minipage}
\caption{\small Histogram of access latencies of seven L1 hits and the 8th being L1 hit or miss when measuring one
target address with pointer chasing (left) on Intel Xeon E5-2690 and (right) on AMD EPYC 7571.}
\label{fig:measure_1set_histogram}
\end{figure}

The major challenge for the receiver is to measure the memory access time precisely and to distinguish an L1 cache hit
and an L1 cache miss (an L2 cache hit or longer).
Table~\ref{table:cache_latency} shows the access latency of L1 hit and L1 miss
on the microarchitectures we tested.
L1 hit takes less than 5 CPU cycles, and L2 hit takes about 10--20 CPU cycles.
Due to the noise caused by the serializing and the granularity of time stamp counter, using \textit{rdtscp} instruction (or \textit{lfence} and \textit{rdtsc} 
instructions) to measure the latency of a memory access cannot distinguish L1 hit from L2 hit. The measurement results of  L1 hit and L2 hit are the same.

Thus, we use  pointer chasing algorithm and a dedicated data structure to measure one memory access precisely.
In the pointer chasing algorithm in Figure~\ref{code:pointer_chasing}, a linked list, where each element stores the address of the next elements, is required.
In the code listed, the $rbx$ points to the head of the linked list.
Since the address of the $mov$ instruction depends on the data fetched from the previous $mov$ instruction,
all the eight accesses are serialized.
However, in a side and covert channel scenario, it is not practical
to use Algorithm~\ref{alg:Protocol1} to build a linked list containing the sender's memory
access destination in a read-only shared library.

Instead of a linked list in the shared library, we use a linked list of 7 elements\footnote{The size of the linked list does not have to be 7. 
However, if the size is small, the noise due to \textit{lfence} will affect the measurements. If the size is large, there will be noise in accessing the elements in the linked list.}
in the
receiver's own memory space, and let the 7$^{th}$ element contain the memory address to be measured.  
In this way, when measuring latency with the pointer chasing algorithm in Figure~\ref{code:pointer_chasing}, it will first access 7 local elements and the target address
at the end. 
Before running the measurement, the receiver can fetch the first 7 local elements to L1 cache,
so the first 7 accesses will always hit in L1 and the total time depends on
whether the 8th element is in L1 cache or not.
To avoid the first 7 elements polluting the LRU state of the target set, the 7 elements can be in one cache set and any other set can be the target set.
Figure~\ref{fig:measure_1set_histogram} shows the result of this measurement strategy (L1 hit of the first 7 elements and the 8th element
being L1 hit or miss). The difference between an L1 hit and an L1 miss of the 8th element is distinguishable on the Intel processors. The latency of L1 hit and L1 miss show different distributions on the AMD processor.

\section{Evaluation}
\label{sec:channel_evaluation}

\begin{algorithm}[t]
\small{
 m: k-bit message to be sent on the channel\\
 $T_s$: sender's sending period\\
 $T_r$: receiver's sampling time\\
 TSC: current time stamp counter, obtained by \textit{rdtscp}
  \algrule
 \textbf{Sender Code:}
 \algrule
\For{$i = 0;\ i <k;\ i = i + 1$}{
\For{an amount time $T_s$}{
  Step 1: Encoding Phase, encoding m[k]
}
}
  \algrule
 \textbf{Receiver Code:}
 \algrule
   \While{True}{
 Step 0: Initializion Phase \\
  \While{TSC $<T_{last}+T_r$}{
 nothing;  
 }
 $T_{last} = $TSC\\
 Step 2: Decoding Phase\\
 }
 }
 \caption{Covert Channel Protocol}
 \label{alg:covert_channel_protocol}
\end{algorithm}

\begin{table}[t]
 \caption{\small Specifications of the tested CPU models.}
 \label{table:CPU_info}
\centering
\small
\begin{threeparttable}
 \begin{tabular}{|p{1.1in}|p{0.5in}|p{0.5in}|p{0.51in}|}
 \hline
\textbf{{Model}} & \textbf{{Intel Xeon E5-2690}}& \textbf{{Intel Xeon E3-1245 v5}} & \textbf{{AMD EPYC 7571}} \\
\hline
Microarchitecture &Sandy Bridge &Skylake &Zen\\
 \hline
Number of cores & 8 & 4 & N/A$^a$\\
 \hline
L1D size of each core&32KB& 32KB & 32KB\\
\hline
L1D associativity& 8-way& 8-way& 8-way\\
\hline
Frequency & 3.8GHz & 3.9GHz  & 2.5GHz\\
\hline
OS & \multicolumn{3}{c|}{16.04.1 Ubuntu} \\
\hline
 \end{tabular}
\begin{tablenotes}
 \small{$^a$We use the AMD processor on Amazon AWS EC2 platform. The CPU model is specific for Amazon AWS. One core was leased for our experiments.}
\end{tablenotes}
\end{threeparttable}
\end{table}

To evaluate the transmission rate of the LRU channel,
we evaluate it as a covert channel using one target set in the L1 data cache.
As shown in Algorithm~\ref{alg:covert_channel_protocol}, 
the sender sends each bit of message $m$ for $T_s$ CPU cycles,
by running the sender's operations (in Algorithm~\ref{alg:Protocol1} or \ref{alg:Protocol2}) for $T_s$ in a loop for each bit in the
message that the sender wants to send.
$T_s$ decides the transmission rate. We calculate the transmission rate with the total number of bits sent divided by the time (measured by $time$ in Linux).
The receiver runs the receiver's operations (in Algorithm~\ref{alg:Protocol1} or \ref{alg:Protocol2}) every $T_r$ CPU cycles in a loop and measures the latency using pointer chasing discussed in Section~\ref{sec:pointer_chasing}. 

The evaluation is conducted on both Intel and AMD processors. The specifications of the tested CPU models are listed in Table~\ref{table:CPU_info}. We evaluated both LRU Channel with shared memory
and without shared memory presented in Section~\ref{sec:LRU_channel} under both hyper-threaded sharing and time-sliced sharing settings.

\subsection{LRU Covert Channels in Intel Processors}

\subsubsection{LRU Channels in Hyper-Threaded Sharing}

For the hyper-threading case,
we tested the covert channel when the sender and the receiver are sharing the same physical core as two hyper-threads.
Each of the sender and the receiver is a process (i.e., a separate program) in Linux.

\begin{figure}[t]
\begin{minipage}{0.49\textwidth}
\centering
\vspace{-12pt}
\includegraphics[width=3.5in]{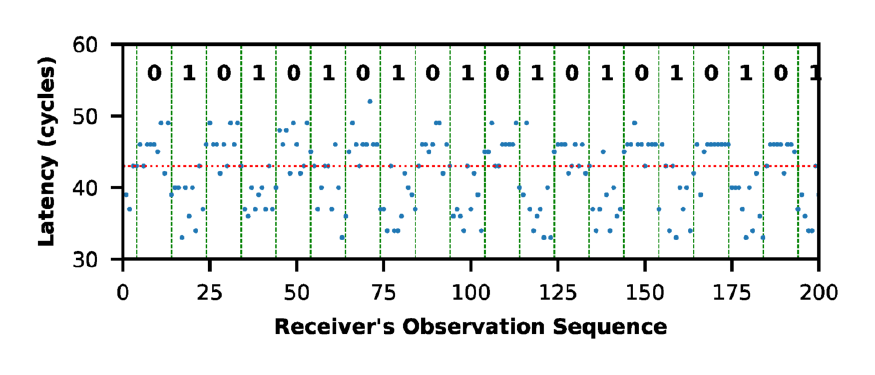}
\end{minipage}
\begin{minipage}{0.49\textwidth}
 \vspace{-20pt}
\centering
\includegraphics[width=3.5in]{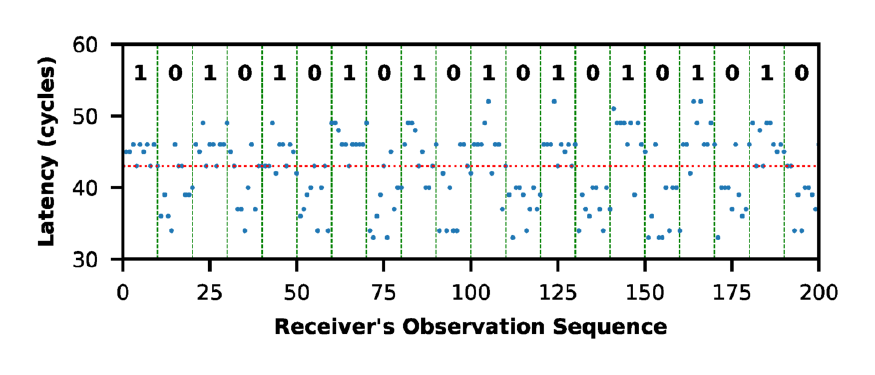}
\end{minipage}
 \vspace{-12pt}
\caption{\small 
Example sequences of receiver's observation when the sender is sending 0 and 1 alternatively on Intel Xeon E5-2690 with a transmission rate of 
480Kbps using (top) Algorithm 1 with $T_r{=}600$, $T_s{=}6000$, and $d{=}8$ and (bottom) Algorithm 2 with $T_r{=}600$, $T_s{=}6000$ and $d{=}4$. The blue dots show the latencies observed by the receiver, and the red dot line shows the threshold of the L1 cache~hit.
}
\label{fig:hyperthread_protocol1_trace}
\end{figure}

\begin{figure*}[t]
\begin{minipage}[t]{0.32\textwidth}
\includegraphics[width=2.3in]{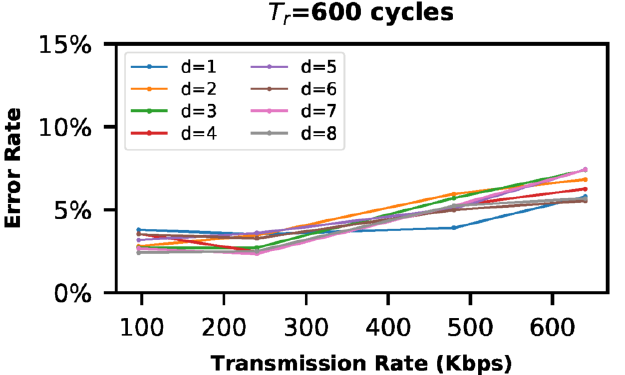}
\vspace{3pt}
\end{minipage}
\enskip
\begin{minipage}[t]{0.32\textwidth}
\includegraphics[width=2.3in]{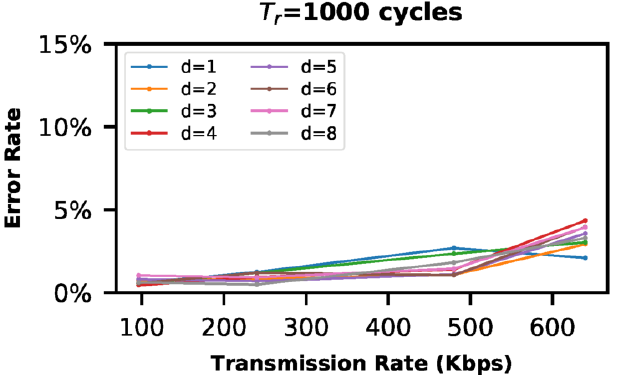}
\end{minipage}
\enskip
\begin{minipage}[t]{0.32\textwidth}
\includegraphics[width=2.3in]{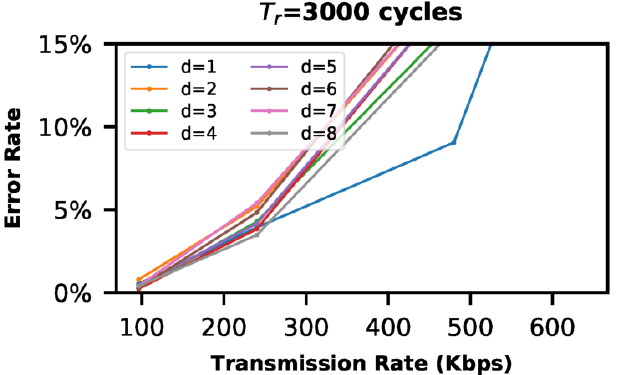}
\end{minipage}
\begin{minipage}[t]{0.32\textwidth}
\includegraphics[width=2.3in]{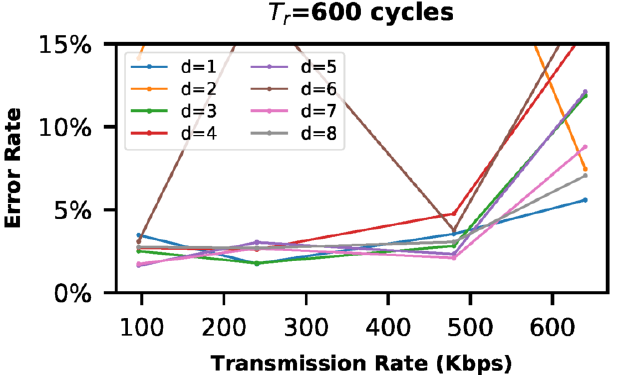}
\end{minipage}
\enskip
\begin{minipage}[t]{0.32\textwidth}
\includegraphics[width=2.3in]{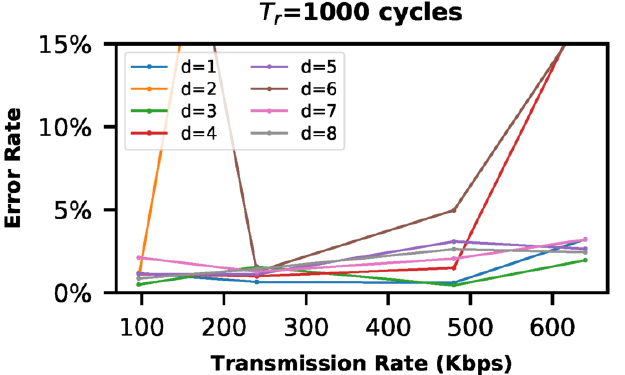}
\end{minipage}
\enskip
\begin{minipage}[t]{0.32\textwidth}
\includegraphics[width=2.3in]{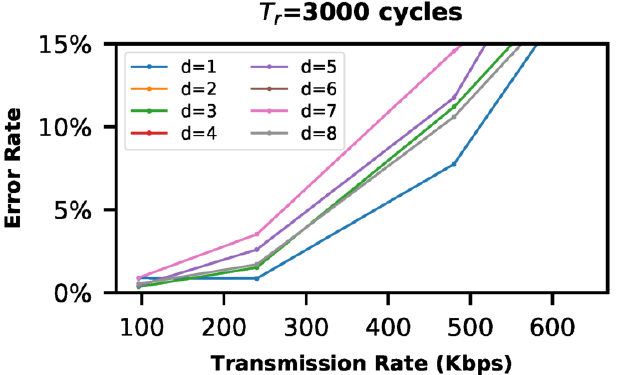}
\end{minipage}
\caption{\small Transmission error rate (evaluated by {\em edit distance}) as a function of the transmission rate (different $T_s$) for different $T_r$ on Intel Xeon E5-2690 using (top) Algorithm 1 and (bottom) Algorithm 2.}
 \label{fig:srv1_ch1_ch2}
\end{figure*}

\textbf{LRU Channel with Shared Memory:}
In Algorithm 1, shared memory is needed among the sender and the receiver processes, e.g., achieved by a shared library.
Figure~\ref{fig:hyperthread_protocol1_trace} (top) shows the traces observed by the receiver
when the sender is sending 0 and 1 alternatively.
When the sender is sending bit 1, the access time of line 0 by the receiver is shorter, as is discussed in Section~\ref{sec:Protocol1}.
Due to the space limit, only the results on Intel Xeon E5-2690 are shown in Figure~\ref{fig:hyperthread_protocol1_trace}.
Evaluation on E3-1245 v5 shows similar results, except that the two processors have different thresholds for L1 hit and miss latencies.
This is due to different latencies for L1 or L2 cache access on the two.
Also, the two processors are running at different frequencies,
and thus, even with the same $T_s=6000$, the transmission rate is 480Kbps for E5-2690 and 580Kbps for E3-1245~v5.

In the evaluation, the sender process sends a random 128-bit binary string repeatedly.
There are 3 types of errors in the channel: 1) bit flips, 2) bit insertions, or 3) bit loss. To evaluate the error rate of the channel, the {\em edit distance}
between the sent string and the received string is calculated using the Wagner-Fischer algorithm~\cite{navarro2001guided}.
We evaluate $T_r=\{600, 1000, 3000\}$ cycles,  and $T_s=\{4500, 6000,12000,30000\}$ cycles.
The receiver's operations of Algorithm 1 in total takes about 560
cycles, including logging of the results,
and thus, $T_r{>}560$.
Because the CPUs have 8-way set-associative caches
and the maximum possible $d$ is $8$, we test parameter $d=\{1,2,3,4,5,6,7,8\}$.
Also, the 128-bit string is sent at least 30 times to obtain the average errors.

Figure~\ref{fig:srv1_ch1_ch2} (top) shows the error rate of the channel versus the different transmission rates (i.e., different values of $T_s$).
As shown in the figure, $d$ does not affect the error rate much on the E5-2690.
This is because, in hyper-threaded sharing, the sender process and the receiver process execute in parallel. 
The sender operation can happen when the receiver is executing any part of his or her operation, and $d$ only makes the sender operation more likely to happen in the sleep part of the receiver's operation.
$T_r=1000$ gives a slightly better error rate than $T_r=600$. This might be because more interleaving between the two threads due to greater $T_r$
and the receiver can observe more sender's activity in one measurement.
As $T_r$ increases to $3000$ cycles, the error rate increases.
In general,  the error rate increases as the transmission rate increases (i.e., $T_s$ decreases). This is because a greater $T_s$ or a smaller $T_r$
will result in  more measurements for each of the bit transmitted, and the noise can be canceled out by taking the average of the measurement~results.

\textbf{LRU Channel without Shared Memory:}
In Algorithm 2, shared memory between the sender and the receiver is not required.
Figure~\ref{fig:hyperthread_protocol1_trace} (bottom)  shows the traces observed by the receiver.
When the sender is sending bit 1, the access time of line 0 by the receiver is longer, due to the sender's access to the same set.

For Algorithm 2, we also evaluate the same set of values of $T_r$, $T_s$, and $d$.
Figure~\ref{fig:srv1_ch1_ch2} (bottom) shows the error rate versus the different transmission rates (different values of $T_s$) on E5-2690. Compared to
LRU channel with shared memory, the LRU channel without shared memory has more noise. As indicated in the simulation result of accessing sequence 2 in Section~\ref{sec:PLRU_analysis}, in Tree-PLRU, when the
sender accesses the set, the receiver may not observe a miss in the end, resulting in a false~0. 
Also, any access to the same set (by the other part of the program or other processes on the core)
may result in a false 1.
However, these errors usually occur consecutively in time. So
the receiver can detect the noise if observing a long sequence of all 1 or all~0. We exclude those traces to obtain
Figure~\ref{fig:srv1_ch1_ch2}.

When $d=\{2,4,6\}$, the error rate is large on E5-2690, especially for large $T_r$. This is because
even $d$ makes the Tree-PLRU point to another side of the sub-tree, and the receiver will not evict line 0 during decoding.

\subsubsection{LRU Channels in Time-Sliced Sharing}
\label{sec:time_slice_intel}

\begin{figure}[t]
\begin{minipage}[t]{0.23\textwidth}
\includegraphics[width=1.6in]{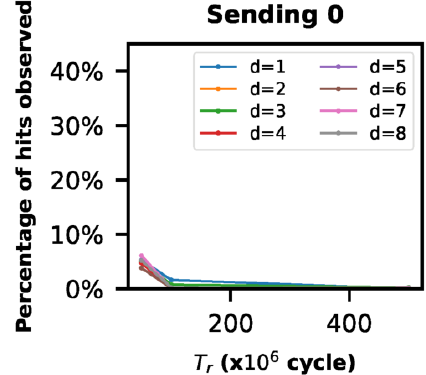}
\vspace{0pt}
\end{minipage}
\begin{minipage}[t]{0.23\textwidth}
\includegraphics[width=1.6in]{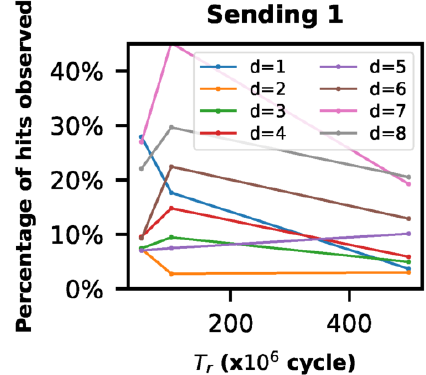}
\end{minipage}
\vspace{-6pt}
\caption{\small Percentage of cache hits observed by the receiver on Intel Xeon E5-2690, when the sender is sending (left) 0 and (right)~1 using Algorithm 1 under time-sliced sharing.}
\label{fig:time_slice_ch1}
\end{figure}

When the sender and receiver are sharing the same core in a time-sliced sharing setting,
the two processes still share the same L1 cache.
To evaluate the covert channel in a time-sliced sharing setting, we programmed the sender process to always send 1 or 0, and the receiver to
measure the time of accessing line 0 every $T_r$.
Figure~\ref{fig:time_slice_ch1} shows the percentage of cache hit received for different $d$ and $T_r$ when
the sender is sending 0 or 1 using Algorithm 1 on both CPUs tested. Each data point comes from 1000 measurements.

As is shown in Figure~\ref{fig:time_slice_ch1}, with proper parameters, the receiver can distinguish
between the sender sending 0 and 1. For example, if $d=8$ and $T_r=10^8$ cycles, the receiver will observe almost 100\% of L1 cache misses when the sender
is sending 0, and the receiver will observe about 30\% of L1 cache hits when the sender is sending 1 on both Intel processors. The receiver does not
observe hits with a higher probability,  because in time-sliced sharing, each process uses the core for a certain period of time. When the receiver
monitors the sender in a loop, multiple loop iterations will run within a time-slice period, and only the first iteration will reflect the sender's
behavior, the other iterations in the time period run without interleaving with the sender. Nevertheless, the receiver can still recognize the message
the sender is sending by the percentage of cache hit received. Assuming 10 measurements are needed when $T_r=10^8$ to differentiate $30\%$ from ${<}5\%$, the transmission rate is about 2.4 bit/s. 

Compared to hyper-threaded sharing, much larger $T_r$ is needed here to have interaction between the two threads (about $10^8$ cycles for
both processors tested).
However, if $T_r$ is too large, the distinguishability decreases, as other processes might be scheduled during $T_r$.
As is shown in Figure~\ref{fig:time_slice_ch1}, $d=8$ and $d=7$ gives the best distinguishability
between the sender sending 0 and 1. This is because $T_r$ is large, and the time for the receiver's operations becomes small compared to the
sleep time. Thus, the context switch is more likely to happen during the sleep time. In Algorithm 1, a greater~$d$ leads to fewer accesses to the target
set after the sleep, and thus, line 0 is less likely to be evicted during decoding.
Such evicted line 0 may result in a false~0.

We also tried to demonstrate Algorithm 2 but failed to observe any signal from the measurement.
We think the reason is that the $T_r$ should be large to allow interference between the sender and the receiver, 
however, any other processes running during $T_r$ could pollute the target set and introduce a lot of noise.

\subsection{LRU Covert Channels in AMD Processors}
\label{sec:channel_evaluation_amd}

For power-savings, AMD L1 cache has a special linear address {\em utag} and {\em way-predictor} (see 2.6.2.2 in \cite{AMD_Family_17h}).
The {\em utag} is a hash of the linear address.
For a load, while the physical address is looked up in TLB, the L1 cache uses the hash of the linear address to match the {\em utag}
and determines which cache way to use in the cache set.
When the physical address is available, only that cache way will be looked up instead of all 8 ways.
So, when the physical address of a load matches a cache line in the cache, if the {\em utag} of that way is of a different linear address
unless the hash of two linear addresses conflicts, a latency of an L1 miss will be observed,
even though the physical address matches and data is in L1.

This makes our Algorithm~\ref{alg:Protocol1} across processes using different address spaces limited.
If the sender process accesses line 0, the {\em utag} of line 0 will be updated with the linear address of line 0 in the sender's address space. When the receiver accesses line 0 and measures the time, unless the hash of the linear address of line 0 in the sender's process and in the receiver's
process conflicts, the receiver will always observe an L1 cache miss latency no matter if the line 0 is in L1 or not.
However, the hash of {\em utag} is not designed for security and is possible to be reverse-engineered.
Furthermore, as long as the sender and the receiver are in the same address space, the LRU channel using Algorithm~\ref{alg:Protocol1} still exists.
For example, it can be used to transfer information in the case of escaping sandbox in JavaScript~\cite{kocher2018spectre}.

\begin{figure}[t]
\begin{minipage}{0.49\textwidth}
\centering
\includegraphics[width=3.5in]{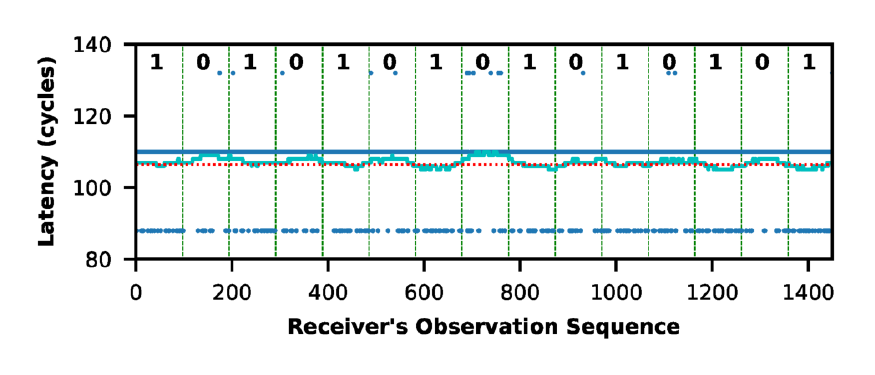}
\end{minipage}
\begin{minipage}{0.49\textwidth}
 \vspace{-20pt}
\centering
\includegraphics[width=3.5in]{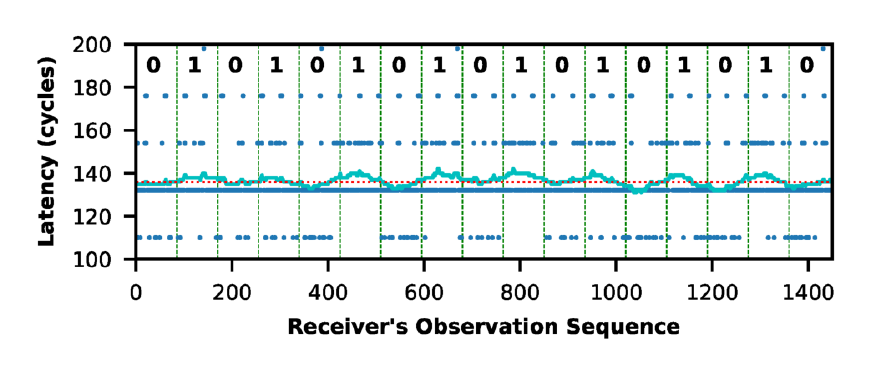}
\end{minipage}
\caption{\small Example sequences of receiver's observation when the sender is sending 0 and 1 alternatively using (top) Algorithm 1 and (bottom) Algorithm 2 on AMD EPYC 7571. For Algorithm 1,  $T_r=1000$, $T_s=10^5$, $d=8$, and the transmission rate is 22Kbps.  For Algorithm 2, $T_r=1000$, $T_s=10^5$, $d=4$, and the transmission rate is 25Kbps. The light blue dot line shows the moving average.}
 \label{fig:hyperthread_amd_trace}
\end{figure}

\begin{figure}[t]
\begin{minipage}[t]{0.23\textwidth}
\includegraphics[width=1.6in]{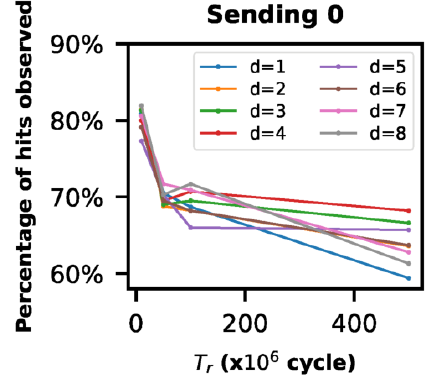}
\end{minipage}
\begin{minipage}[t]{0.23\textwidth}
\includegraphics[width=1.6in]{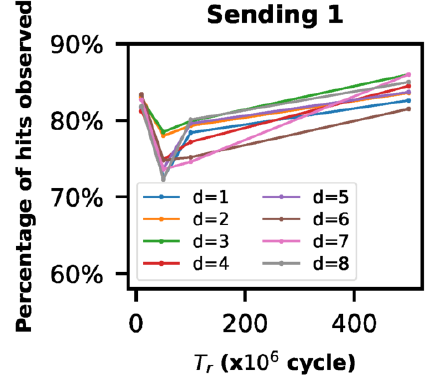}
\end{minipage}
\caption{\small Percentage of Percentage of cache hits observed by the receiver on AMD EPYC 7571,
when the sender and receiver are sharing a core in a time-slice setting and the sender is sending (left) 0 and (right) 1 using Algorithm 1.}
 \label{fig:AMD_time_slice_ch1}
\end{figure}

We evaluate the characteristics of the LRU covert channel on AMD EPYC 7571 processor on Amazon AWS EC2 platform.
Figure~\ref{fig:hyperthread_amd_trace} (top) shows the trace observed by the receiver, when the receiver and the sender are two threads in the same address space (using $pthread$s in C)
running in a hyper-threaded sharing using Algorithm~\ref{alg:Protocol1}.
Due to the coarse granularity of the readout value of the time stamp counter in AMD, it is hard to identify the signal from the raw measurements (blue dots).
The light blue dot line in Figure~\ref{fig:hyperthread_amd_trace} shows the moving average of the latency of 97 measurements, where the 97 is the best fit period of sending one bit for this
trace\footnote{The fact that the period does not equal to $T_s/T_r$ indicates that threads do not get scheduled evenly. This might be due to the Amazon EC2 platform, as we observe similar phenomenon on Intel processors on EC2.}.
When the sender is sending 0 and 1 alternatively, the moving average is a wave-like pattern, meaning the receiver can receive the message from the sender.
By measuring the total time taken by the receiver to gather the trace and the period of each bit received, the effective transmission rate is 22Kbps.
Due to the coarser-granularity of the AMD time stamp counter and lower frequency,
the transmission rate of the channel is about one order of magnitude lower than that in Intel processors.

We also tested Algorithm~\ref{alg:Protocol2} under hyper-threaded sharing on AMD EPYC 7571.
Figure~\ref{fig:hyperthread_amd_trace} (bottom) shows a trace observed by the receiver.
The receiver and the sender are two programs (in different memory space).
Similarly, the light blue dot line shows the moving average of the latency of 85 measurements, where the 85 is the best fit, resulting in an effective transmission rate of 25Kbps.
When the sender is sending 0 and 1 alternatively, the moving average is a wave-like pattern, meaning the receiver can receive the message from the sender.
The measured latency in Figure~\ref{fig:hyperthread_amd_trace} (top) and (bottom) are quite different.
This might due to the processor running at a different frequency for power saving at the time of measurement.

We further tested Algorithm~\ref{alg:Protocol1} under time-sliced sharing setting using $pthread$s.
Figure~\ref{fig:AMD_time_slice_ch1} shows the different results observed by the receiver when the sender is sending 0 and 1.
The thresholds to decide whether a latency represents hit and miss are selected such as to maximize the difference between 0 and~1.
As shown in Figure~\ref{fig:AMD_time_slice_ch1}, when $T_r=10^8$
cycles, the receiver will receive about $70\%$ of L1 cache hits when the sender is sending 0, and about  $77\%$ of L1 cache hits when the sender is sending 1.
This is enough to differentiate 0 and 1, by examining if percentage of cache hit is below or above the threshold.
The transmission rate is about 0.2 bits per second.
When increasing $T_r$, more interleaving between the sender thread and the receiver thread happens during each measurement taken by the receiver,
and the difference between 0 and 1 gets greater indicating less noise.  
The parameter $d$ does not play a significant role.
We do not observe any signal using Algorithm~\ref{alg:Protocol2} in time-sliced sharing, similar to the case for Intel.

\subsection{Comparing the Evaluated LRU Channels}

Table~\ref{table:trans_rate_sum} compares the transmission rate per cache set of the channels tested with different configurations.
Hyper-threading gives a much higher transmission rate than time-sliced sharing because of more interference between the sender and the receiver.
Under hyper-threading, Algorithm~\ref{alg:Protocol1} and Algorithm~\ref{alg:Protocol2} have similar transmission rate.
The transmission rate is comparable to other timing channels in caches~\cite{liu2015last,yao2018coherence}.
However, recall that Algorithm~\ref{alg:Protocol2} is easily affected by noise due to activities of other programs,
but the noise is easy to filter, because the noise activity is usually of a different frequency.
The LRU channel on AMD processors is about one order of magnitude slower than on Intel processors,
due to the coarser-granularity of readout value of timestamp counter and lower clock frequency.

\begin{table}[t]
\caption{\small Transmission rate of the evaluated LRU channels.}
\label{table:trans_rate_sum}
\centering
\small
\begin{tabular}{|c|c|c|c|}
\hline
&  & \textbf{Intel} & \textbf{AMD} \\
\hline
\multirow{ 2}{*}{\textbf{Hyper-Threaded}}& {Algorithm 1}& $\sim$500Kbps &  $\sim$20Kbps\\
\cline{2-4}
& {Algorithm 2}& $\sim$500Kbps & $\sim$20Kbps \\
\hline
\multirow{ 2}{*}{\textbf{Time-Sliced}} & {Algorithm 1}&  $\sim$2bps & $\sim$0.2bps \\
\cline{2-4}
& {Algorithm 2}&  --  & -- \\
\hline
\end{tabular}
\end{table}

\section{Stealthiness of LRU Channels}
\label{sec:comp}

\begin{table*}[t]
\begin{minipage}{0.33\textwidth}
\centering
\caption{\small Latency of Encoding (cycles).}
\label{table:encoding_time}
\small
\begin{tabular}{|p{0.73in}|p{0.35in}|p{0.25in}|p{0.5in}|}
\hline
&  \textbf{F+R (mem)} & \textbf{F+R (L1)} & \textbf{L1 LRU (Alg.1\&2)} \\
\hline
\textbf{{Intel Xeon E5-2690}} & 336 & 35 &  31\\
\hline
\textbf{{Intel Xeon E3-1245 v5}}  & 288 & 40& 35 \\
\hline
\textbf{{AMD EPYC 7571}} & 232 & 56  & 52 \\
\hline
\end{tabular}
\end{minipage}
\quad \quad
\begin{minipage}{0.6\textwidth}
\caption{\small Cache Miss Rate of the Sender Process.}
\label{table:cache_miss}
\centering
\small
\begin{tabular}{|p{0.65in}|p{0.24in}|p{0.35in}|p{0.25in}|p{0.48in}|p{0.48in}|p{0.35in}|p{0.35in}|}
\hline
&& \textbf{{F+R (mem)}}  & \textbf{{F+R (L1)}} & \textbf{{L1 LRU Alg.1}} & \textbf{{L1 LRU Alg.2}} &\textbf{{sender \& gcc}} &\textbf{{sender only}}  \\
\hline
\multirow{3}{0.7in}{\textbf{{Intel Xeon E5-2690}}} &  L1D &0.07\% & 0.04\% & 0.03\% & 0.03\% & 0.03\% & 0.01\%\\
\cline{2-8}
& L2& 62\% & 6.67\% & 9.59\% & 15.6\% &  31\%  & 8.32\%\\
\cline{2-8}
& LLC& 88\% & 0.77\% & 0.71\% & 1.07\% &  61\%  & 1.46\%\\
\hline
\multirow{3}{0.7in}{\textbf{{Intel Xeon E3-1245 v5}}} &  L1D &0.06\% & 0.02\% & 0.01\% & 0.01\% & 0.01\%   & 0.00\% \\
\cline{2-8}
& L2& 63\% & 11\% & 17\% & 14\%  & 48\% & 26\% \\
\cline{2-8}
& LLC& 92\% & 8.12\% & 8.15\% & 7.42\%  & 70\% & 27\% \\
\hline
\end{tabular}
\end{minipage}
\end{table*}

In most of the existing cache side channels, the receiver measures whether certain cache line exists in the cache directly. 
For example, in the Flush+Reload attack~\cite{yarom2014flush}, the sender fetches a cache line into the cache,
and the receiver measures directly whether a certain cache line is in the cache.
To build a channel, the cache replacement should happen due to the sender's access.
Meanwhile, in our LRU cache channel, the sender's operations does not need to cause any cache replacements,
because the LRU states are updated on both cache hits and misses.
Instead, the cache replacement happens when the receiver wants to measure the LRU state during the decoding phase.
This makes the LRU channel more stealthy on the sender's side.

Table~\ref{table:encoding_time} shows the encoding time of the sender. 
The encoding times in the table include the time to calculate the victim address. 
For LRU channels, it is assumed that the victim line is already in the cache before the attack.
The LRU channels are compared with the Flush+Reload channels. 
We implemented two variants, the one denoted by F+R (mem) uses {\em clflush} instruction to flush the data all the way down to memory, 
while the on denoted by F+R (L1) uses eight accesses to the L1 cache set to evict the data from L1.
As is shown in the table, both LRU channels require less encoding time than F+R channels. 
Because for the LRU channels, the sender can encode the message with cache hits, 
while the Flush+Reload channels always require the sender to have cache misses in the target cache level.

Table~\ref{table:cache_miss} shows the cache miss rate of the sender process measured using Linux Perf tool from hardware performance 
counters\footnote{We do not have access to the hardware performance counter on AMD machines on Amazon AWS,  
so only result from local Intel machines are shown in Table~\ref{table:cache_miss}.}.  The results show that the sender of 
LRU Algorithm~\ref{alg:Protocol1} and Algorithm~\ref{alg:Protocol2} have smaller L1 cache miss rate than the Flush+Reload. 
To provide a baseline of no attack, we also show the results when there is only the sender process running on the physical core (denoted by {\em sender only}) 
and the results with the sender sharing the physical core with a benign $gcc$ workload (denoted by {\em sender} \& {\em gcc}). 
When there is only the sender process, it has the smallest L1 miss rate\footnote{The {\em sender only} case still has a relatively high L2 and LLC miss rate due to fewer references to the L2 and LLC.}.
When it is sharing the core with a benign program, the benign program, e.g., the $gcc$,
will cause contention in the cache, similar or even bigger to the contention due to the receiver in the LRU channel.
Hence, if a victim wants to detect a potential cache side channel attacks using performance counters~\cite{zhang2016cloudradar,chiappetta2016real,alam2017performance}, the LRU channel is difficult to detect
as it may not be distinguished from the contention due to benign programs.

\section{LRU Channels in Transient Execution Attacks}
\label{sec:spectre_LRU}

\begin{table}[t]
\caption{\small Cache Miss Rate of Spectre V1 Attack.}
\label{table:cache_miss_spec}
\centering
\small
\begin{tabular}{|p{0.65in}|p{0.24in}|p{0.35in}|p{0.25in}|p{0.48in}|p{0.48in}|}
\hline
&& \textbf{{F+R (mem)}}  & \textbf{{F+R (L1)}} & \textbf{{L1 LRU Alg.1}} & \textbf{{L1 LRU Alg.2}}\\
\hline
\multirow{ 3}{0.65in}{\textbf{{Intel Xeon E5-2690}}} &  L1D &2.75\% & 4.73\% & 4.19\% & 4.75\%  \\
\cline{2-6}
& L2& 7.58\% & 0.07\% & 0.11\% & 0.09\% \\
\cline{2-6}
& LLC& 98.15\% & 0.87\% & 0.72\% & 0.87\% \\
\hline
\multirow{ 3}{0.65in}{\textbf{{Intel Xeon E3-1245 v5}}} &  L1D &2.86\% & 4.84\% & 4.13\% & 4.86\%  \\
\cline{2-6}
& L2& 7.39\% & 0.49\% & 0.71\% & 0.45\% \\
\cline{2-6}
& LLC& 91.17\% & 1.83\% & 0.74\% & 0.96\% \\
\hline
\end{tabular}
\end{table}

Transient execution attacks, e.g., Spectre, leverage transient execution to access secret and a covert channel to pass the secret to the attacker~\cite{kocher2018spectre,lipp2018meltdown,canella2018systematic}.
Currently, most proof-of-concept codes  of transient execution attacks use the cache Flush+Reload covert channel.
Here we demonstrate that our LRU covert channel also works with Spectre to retrieve the secret.

Note that here the secret contains more than 1 bit, and multiple cache sets are used to encode the secret.
In practice, 63 cache sets are used (both Intel and AMD processors tested have 64 sets, remaining one set is for the 
7 elements in the pointer chasing algorithm as discussed in Section~\ref{sec:measure_L1}). 

The Flush+Reload covert channel needs one memory access depending on the secret as the sender's operation. 
Meanwhile, as shown in both algorithms in Section~\ref{sec:LRU_channel}, the sender's operation in the LRU channels also only need one memory access whose target set depends on the secret.
Thus, the victim code using the LRU channel can be identical to the disclosure gadget in the Flush+Reload channel.
Thus, when demonstrating transient execution attack using the LRU channels, 
we take the Spectre variant 1 attack sample code~\cite{kocher2018spectre} and keep the victim (sender) code to be the same, and change the attacker
(receiver) code to use the L1 LRU channels as the disclosure primitive instead.
We are able to launch the Spectre attack using the LRU channels (both Algorithm 1 and 2) to observe the secret.
Also, Table~\ref{table:cache_miss_spec} shows the cache miss rate (including both the victim and the attacker) during a Spectre attack. 

Comparing to the Flush+Reload channel, the advantage of the LRU disclosure primitive is the short encoding time (i.e., the sender's operations), 
and thus, a smaller speculative window is required, which may make the attack more dangerous and harder to defend.

\section{LRU Attack and Secure Caches}
\label{sec:lru_and_sec_caches}

\begin{figure}[t]
\centering
\includegraphics[width=3.3in]{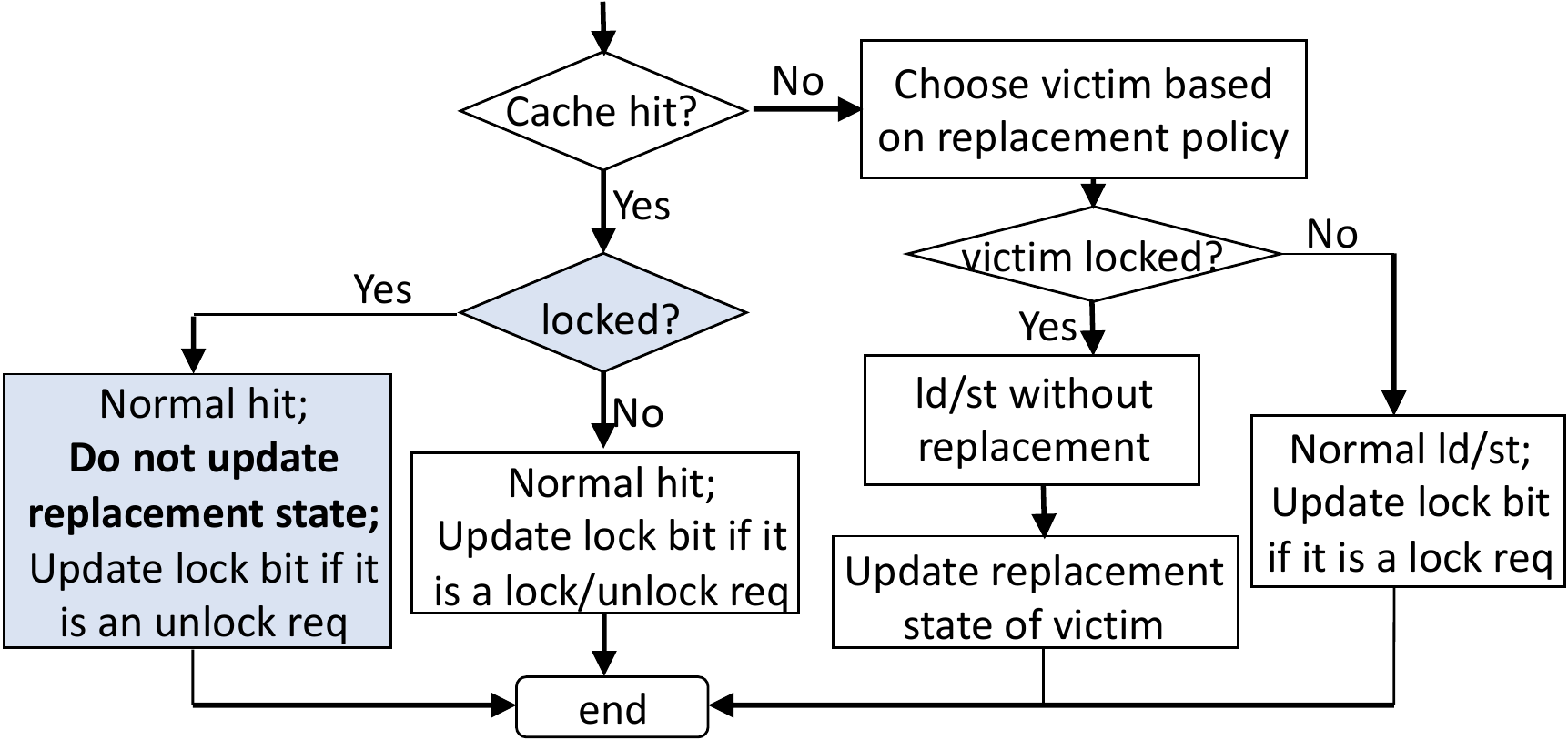}
 \caption{\small PL cache replacement logic flow-chart. White boxes show the original PL cache design in~\cite{wang2007new}. 
 Blue boxes show the new PL logic added in our simulation to defend the LRU~attack.}
 \label{fig:PL_cache}
\end{figure}

\begin{figure}
\centering
\includegraphics[width=3.3in]{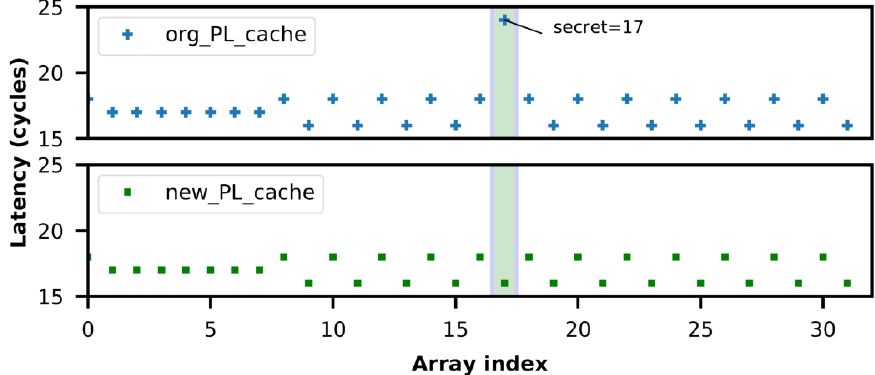}
 \caption{\small Simulation result of the LRU attack Algorithm 2 in {\tt gem5} with (top) original PL cache design and (bottom) new PL cache design which locks the LRU state to defend the LRU~attack.}
 \label{fig:PL_attack}
\end{figure}

\begin{figure*}[t]
\begin{minipage}[t]{\textwidth}
\centering
\includegraphics[width=6in]{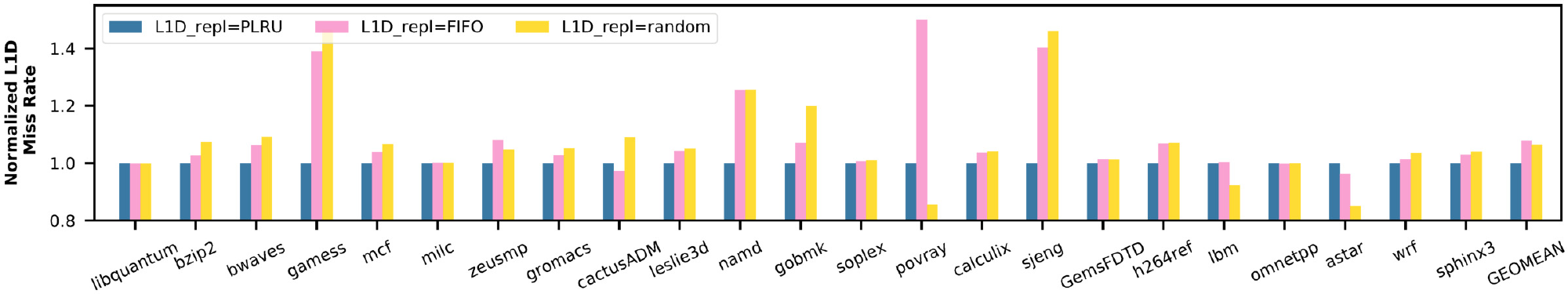}
\end{minipage}
\begin{minipage}[t]{\textwidth}
\centering
\includegraphics[width=6in]{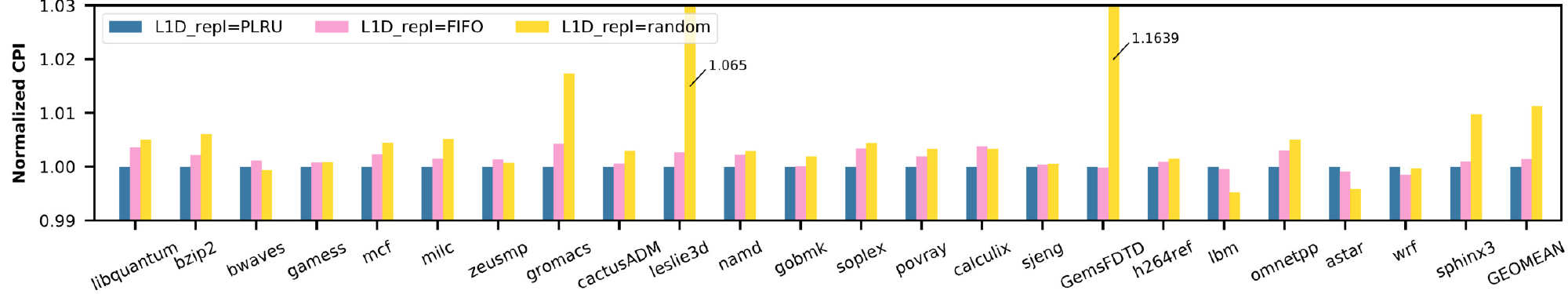}
 \end{minipage}
 \caption{\small (top) Cache miss rate of L1 Data cache and (bottom) normalized CPI when different cache replacement policies (Tree-PLRU, FIFO, random) are used in the L1~Data~cache.
The results are normalized with the result of Tree-PLRU policy.}
 \label{fig:l1d_gem5}
\end{figure*}

Several designs have been proposed to defend the conventional and transient execution attacks, using partitioning or randomization.
Some defenses of transient execution attacks that stop the transient execution but leave the covert channel open, such as~\cite{taram2019context}, are not the focus of this paper. 

\textbf{Partitioning:}
Many secure caches partition the cache (tag and data) between the victim and the
attacker~\cite{lee2005architecture,wang2007new,domnitser2012non,zhang2012language,costan2016sanctum},
but the replacement policy is not considered or specified. 

For example, in Partition-Locked (PL) cache~\cite{wang2007new}, each cache line is extended with one lock bit.
When a cache line is locked, the line will not be evicted by any cache replacement until unlocking to protect the line, as shown in Figure~\ref{fig:PL_cache}. 
If a locked line is chosen as victim to be replaced, the replacement will not happen, and the incoming line will be handled uncached. It is shown that the PL cache can effectively defend Flush+Reload, Prime+Probe, and other attacks.

But the LRU state will still be updated on accesses to the locked cache line, and the update  will affect the LRU states of other lines.
We implemented the PL cache using PLRU replacement algorithm in the {\tt gem5} simulator, and tested the LRU attack. 
During the test, line N (the line accessed by the sender) is first locked by the sender, and Algorithm~\ref{alg:Protocol2} is used to build a channel\footnote{Algorithm~\ref{alg:Protocol1} is 
protected by PL cache when line 0 is locked. Because line 0 will not be evicted in the decoding phase, and the receiver will always get 
a cache hit no matter what the sender is sending.}. As shown in Figure~\ref{fig:PL_attack} (top), 
with the original design, the receiver can still receive the secret by observing the time of accessing line 0.
This is because the sender's access to the locked line will change the eviction order of lines that are not locked, which can be observed by the receiver later.
To mitigate the LRU channel, the LRU state should be locked as well.  We add the blue boxes in Figure~\ref{fig:PL_cache} to PL cache design. In this way, 
the receiver will always observe a cache hit, and thus not learn any information, as shown in Figure~\ref{fig:PL_attack} (bottom).

Other work, such as DAWG~\cite{kiriansky2018dawg}, also
proposes to partition the cache and the PLRU states  in a cache set between protection domains.
And the LRU channel can be mitigated.
We are unaware of any other designs that partition the LRU states, and secure cache designers need to be careful to consider LRU based attacks.

To mitigate transient execution cache side-channel attacks, InvisiSpec~\cite{yan2018invisispec} 
proposes to only update micro-architectural states (including the LRU state) after the access is not speculative. So the LRU channels cannot be used in transient execution attacks, if InvisiSpec defense is applied.

\textbf{Randomization:}
Other secure cache designs use randomization.
For example, Random fill cache~\cite{liu2014random} decouple the access and the cache line brought into the cache, by fetching 
a random cache line instead of the cache line being accessed. However, if the cache line is already in the cache, 
on a cache hit, the replacement state will still be updated, and the LRU channel could still work.
Meanwhile, some designs randomize the mapping between the addresses and the cache sets, such as New cache, RP cache, or
CEASER cache~\cite{wang2007new,liu2016newcache,qureshi2018ceaser}. So the receiver (and the sender) cannot map the addresses to the target cache set to build a channel.

\section{Defending the LRU Channels}
\label{sec:defence}

The LRU timing-based channels leverage the fact that the sender and the receiver share the LRU states in caches.
Thus, there could be several approaches to defend the LRU timing-based channels.
Other than the secure caches mentioned in Section~\ref{sec:lru_and_sec_caches}, another mitigation is to use another cache replacement policy instead of LRU or PLRU.
In this way, no more LRU state exists, and the channel is removed.

\textbf{Random Replacement Policy:}
Random replacement policy does not need any states in the cache.
Every time a replacement is needed, a random cache way in the cache set will be evicted.
For simplicity, most ARM processors use a pseudo-random replacement policy~\cite{ARM_replace}, and naturally defend the LRU attack.

\textbf{FIFO Replacement Policy:}
First-In First-Out (or Round-Robin) replacement policy selects the oldest cache line that is fetched into the cache to be the victim. 
States are still required to store the history of cache lines fetched into cache. 
And thus, FIFO state still contains extra information than which cache line
is presented in the cache.
However, different from LRU, the FIFO states are only
updated when a new cache line is brought into the cache on cache misses. 
Thus, it would require the sender to trigger a cache miss to let the FIFO state be able to be observed by the receiver, similar to the  existing cache channels.

\textbf{Performance Evaluation of Random and FIFO Policies:}
LRU replacement policy is widely used in processors because of its performance.
In this section, we evaluate the performance of different replacement policies in the {\tt gem5} simulator~\cite{binkert2011gem5}.
We simulated a single out-of-order CPU core and a memory system with 2-level caches (32KiB 4-way L1I, 64KiB 8-way L1D with a latency
of  4 cycles, 2MiB 16-way L2 with a latency of 8 cycles, and main memory latency of 50 ns).
SPEC 2006 int and float benchmarks were tested~\cite{henning2006spec}.
Since we focus on the LRU channels in the L1 data cache,
we tested different replacement policies in L1 data cache.

As shown in Figure~\ref{fig:l1d_gem5} (top), compared to Tree-PLRU, the FIFO and Random replacement policies give small degradation
on L1 data cache miss rate overall. Depending on the benchmark, FIFO and Random replacement policy sometimes have an even lower
cache miss rate than Tree-PLRU.
Since an L1 miss can still hit in L2, the overall CPU performance, indicated by cycles per instruction (CPI) in Figure~\ref{fig:l1d_gem5} (bottom), is
only changed less than 2\% compared to the baseline. Thus, using a different replacement policy in the L1 data cache to mitigate the LRU side and
covert channel only gives small overhead -- while increasing security.  Similarly, if the channels in all the levels of cache are
to be mitigated, the replacement policies of all the levels of caches need to be changed.

\section{Conclusion}
\label{sec:conclusion}

We presented novel timing-based channels leveraging the cache LRU replacement states.
We designed two protocols to transfer information between processes using the LRU states for both cases 
when there is shared memory between the sender and the receiver and when there is no shared memory.
We also demonstrated the LRU channels on real-world commercial processors.
The LRU channels require access (cache hit or miss) from the sender,
while all the existing state-based timing-based cache side and covert channels always need the sender to trigger a cache replacement (a cache miss).
Thus, the LRU channel has shorter encoding time, lower cache miss rate for the sender,
and requires a smaller speculation window in transient attack scenarios.
We show the new LRU channels also affect the current secure cache designs.
In the end, we proposed several methods to mitigate the LRU channel and evaluated them,
including a modified design of a secure PL cache.

\section*{Acknowledgement}
\label{sec:acknowledgment}

We would like to thank the authors of InvisiSpec~\cite{yan2018invisispec}, especially Mengjia Yan, for their open-source code and scripts. 
Special thanks to Linbo Shao and Junwen Shao for helping with {\tt gem5} simulation.
We would like to acknowledge Amazon for providing AWS Cloud Credits for Research.
This work was supported by NSF 1651945 and 1813797, and through SRC award number
2844.001.


\bibliographystyle{ieeetr}
\bibliography{ref}

\end{document}